# Optical Magnetic Response in a Single Metal Nanobrick

Jianwei Tang, Sailing He, et al.

**Abstract:** Anti-symmetric localized surface plasmons are demonstrated on a single silver nanostrip sandwiched by SiC layers. By employing the resonance of anti-symmetric localized surface plasmons, we enable single metal nanobricks to produce optical magnetism, in the blue and violet light range, as well as in a part of the ultraviolet light range. The physical mechanism is explained.

**T**he magnetic response of a natural material is usually very weak, especially at optical frequencies, where people simply put magnetic permeability $\mu = 1$. However, with the help of metamaterials with artificial structures on sub-wavelength scales,, people have managed to realize strong magnetic response from microwave frequencies to optical frequencies. In common designs of metamaterials with optical magnetic response, the magnetism relies on the excitation of an anti-symmetric plasmon resonance of a pair of coupled metal nanostrips (or other similar geometries, e.g. nanopillars, nanorods, etc.) [1, 2]. The anti-symmetric plasmon resonance arises due to the coupling of the localized surface plasmons of neighboring nanostrips within a pair. In general it is observed that only one fundamental plasmon resonance exists for the localized surface plasmon of a single metal nanostrip, which bjehaves like an electric dipole. When two metal nanostrips are positioned closely to each other, the fundamental plasmon resonance splits into two resonances, referred to as the symmetric and anti-symmetric resonances. The anti-symmetric plasmon resonance behaves like a magnetic dipole and thus produces magnetism. From all the relevant works reported so far, one may conclude that pairing metal nanostrips is necessary for a magnetic response. Such a conclusion is not true. In fact, with an appropriate design, a single metal nanostrip is also able to produce magnetism. In this work, we demonstrate anti-symmetric localized surface plasmons on a single metal nanostrip. By employing the resonance of such an anti-symmetric localized surface plasmon, we enable a single metal nanobrick to produce optical magnetism, in the blue and violet light range, as well as in a part of the ultraviolet light range.

To demonstrate anti-symmetric localized surface plasmons on a single metal nanostrip, we sandwich every silver nanostrip (infinitely long as a 2D case) between two SiC nanostrips and let a plane wave with the electric field polarized in the x direction illuminate normally from the top (Fig. 1).

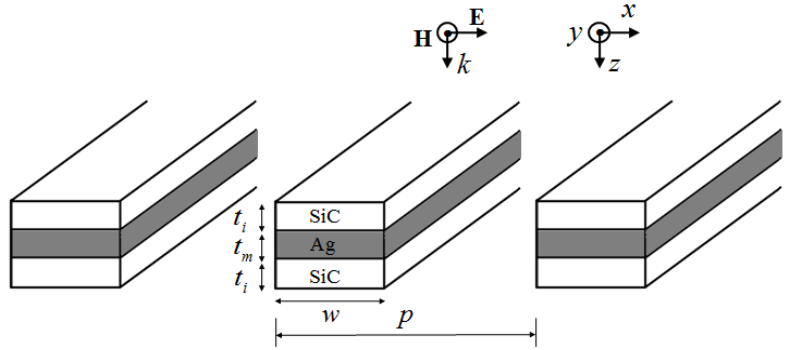

FIG. 1 Silver nanostrips sandwiched by SiC nanostrips and illuminated normally from the top by a plane wave with the electric field polarized in the x direction. The thickness of the silver layer is $t_m = 15$ nm and the thickness of the SiC layer is $t_i = 30 nm$. The width of the nanostrips is $w = 70 nm$. The lattice constant of the periodic nanostrip array is $p = 300 nm$.

When the nanostrips are illuminated by the incident electromagnetic wave, localized surface plasmons are excited on the two Ag/SiC interfaces. The localized surface plasmons on the two interfaces interact with each other through the fields penetrated into the metal, which causes the plasmon resonance to split into symmetric resonance (Fig. 2(b)) and anti-symmetric resonance (Fig. 2(c)) as shown by the red transmission spectrum line in Fig. 2(a). For the symmetric resonance, the electric field on the two surfaces of the silver nanostrip is in phase. This resonance is simply the kind of localized surface plasmon resonance commonly observed for a single metal nanostrip. For the anti-symmmetric resonance, the electric field on the two surfaces is out of phase. To the best of our knowledge, such an anti-symmetric localized surface plasmon on a single metal nanostrip has never been reported before. This may be due to the fact that the SiC layers are essential in order to achieve the anti-symmetric resonance of our single silver nanostrip, as will be analyzed in detail later. If the strip width $w$ is large (e.g., 140 nm), the high order anti-symmetric localized surface plasmon resonant modes will be revealed in the transmission spectrum. In the transmission spectrum we also find that the anti-symmetric resonance is at a higher frequency than the symmetric resonance. This is contrary to the case of a coupled pair of metal nanostrips, where the anti-symmetric resonance is at a lower frequency than the symmetric resonance. The key difference of the underlying mechanism between these two cases is that for a single metal nanostrip the localized surface plasmons on the two surfaces couple with each other through the field inside the metal medium, while for a coupled pair of metal nanostrips, the localized surface plasmons couple in the dielectric medium.

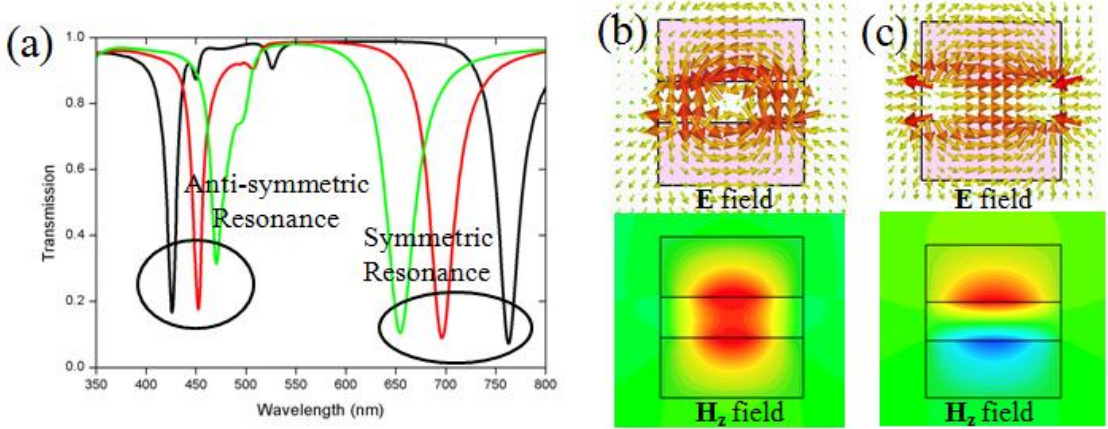

FIG. 2 (a) The transmission spectra of our samples. For all the samples, $p$ = 300 nm, $t_i$ = 30 nm, and $w$ = 70 nm. The thickness of the silver layer is $t_m$ =15 nm (black line), 20 nm (red line), 30 nm (green line) for the three samples. (b) The electric field and $H_z$ field distributions of the anti-symmetric plasmon resonance. (c) The electric field and $H_z$ field distributions of the symmetric plasmon resonance. For all the numerical simulation in this paper, the permittivity of silver is obtained from experimental data [3] and SiC is approximately described as dielectric medium with constant permittivity of 8.

To investigate further the physical principle of the sandwiched silver nanostrip, we treat it as a 2-D IMI (insulator/metal/insulator) waveguide of finite length $w$ as shown in Fig. 3(a) [4]. The thickness of the Ag layer is $t_m$, the thickness of the SiC layer is $t_i$. The blue lines in Fig. 3(b) are the analytical dispersion curves for the two fundamental propagating modes of the IMI waveguide of infinite length. For simplicity, here the imaginary part of the permittivity of silver [3] is neglected. The upper dispersion curve corresponds to the odd propagating mode ($E_x(z)$ is an odd function and $H_y(z)$ is even, as shown in Fig. 3(c)), and the lower curve corresponds to the even propagating mode ($E_x(z)$ is an even function and $H_y(z)$ is odd, as shown in Fig. 3(d)). These two fundamental propagating modes in the IMI waveguide are due to the evanescent field coupling of the propagating surface plasmon polaritons at the two Ag/SiC interfaces. If the IMI waveguide is of finite length, reflections at the two terminations occur and the interference of the counter-propagating guided waves results in some Fabry–Pérot (F-P) resonances with standing-wave-like field distribution. Such a field distribution is similar to the field distribution of the plasmon resonance of our sandwiched single silver nanostrip (studied earlier in Fig. 2): F-P resonance of the even propagating mode is similar to the symmetric localized surface plasmon resonance (see Fig. 2(c)), and F-P resonance of the odd mode is similar to the anti-symmetric resonance (see Fig. 2(b)). In this sense, the localized surface plasmon resonances of a single metal nanostrip can be viewed as F-P resonances of counter-propagating guided waves, which are partially reflected at the waveguide terminations [4].

Two requirements should be fulfilled for the waveguide to have a strong F-P resonance: first, the effective wavelength of the guided wave should match the length of the waveguide (i.e., the length of the waveguide should be an integer of half effective wavelength); secondly, the reflections at the two terminations should be strong enough for the guided wave. The SiC cladding layers make the dispersion curves of the two fundamental guided modes lower than the light line in SiC (solid black line in Fig. 3(b)), and thus far below the light line in air (dashed black line in Fig. 3(b)). Otherwise, the dispersion curves (purple lines in Fig. 3(b) for the silver waveguide without SiC cladding layers) will be close to the light line in air. Thus, with the SiC cladding layers the effective wavelength of the guided waves can be much shorter than that in air, which makes it possible to support F-P resonance in a waveguide of sub-wavelength length (required for the homogenization of the metamaterial). With such a short effective wavelength, the fields are tightly confined inside the SiC/Ag/SiC waveguide structure (see Fig. 3(c) and (d)). This guarantees the large reflection of the guided waves at the waveguide terminations. Although the SiC layers play roles for both symmetric and anti-symmetric localized surface plasmon resonances, we emphasize here that they are essential for the anti-symmetric resonances, but not necessary for the symmetric resonances.

The silver thickness $t_m$ is an important parameter of the waveguide, as it is closely related to the coupling between the surface plasmons at the two SiC/Ag interfaces. For a smaller $t_m$ (e.g., 15 nm), the dispersion curves (red lines in Fig. 3(b)) of the even and odd modes are more separated. For a larger $t_m$ (e.g., 25 nm), the dispersion curves (green lines in Fig. 3(b)) are closer to each other, until the two modes degenerate for a large enough $t_m$ when the surface plasmons at the two interfaces become decoupled [5]. We have also plotted the transmission spectra for different $t_m$ (15 nm, 20 nm, and 25 nm) in Fig. 2(a). For a smaller $t_m$, the anti-symmetric resonances blue shift and the symmetric resonances red shift; while for a larger $t_m$, the anti-symmetric resonances blue shift and the symmetric resonances red shift. This agrees with the dispersion curves shown in Fig. 3(b).

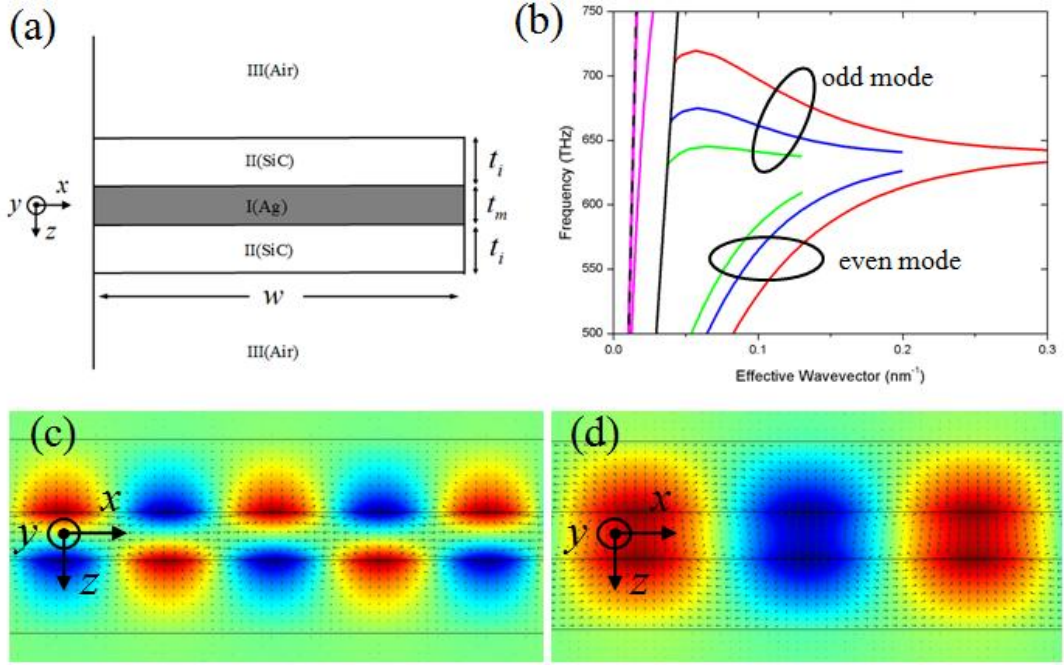

FIG. 3 (a) 2-D IMI (insulator/metal/insulator) waveguide of finite length $w$. Wave is guided in the x direction. The thickness of SiC layers is $t_i = 30$ nm. (b) Dispersion curves of the waveguides. The red, blue and green lines are the dispersion curves of the waveguide with Ag layer thickness $t_m =$ 15 nm, 20 nm and 25 nm, respectively. The solid and dashed black lines are light lines in SiC and air, respectively. The purple lines are the dispersion curves of the silver waveguide without SiC cladding layers. The electric field distribution (arrow plot) and magnetic field distribution (color contour plot) of (c) the even mode and (d) the odd mode.

When we treat the localized surface plasmon resonances of the single metal nanostrip as F-P resonances of guided waves, the way of excitation/illumination is not involved in such a modal analysis. However, we know the way of excitation/illumination would influence somehow the localized surface plasmon resonance. First, the strength of a certain resonant mode depends on its capability of scattering the exciting/illuminating light. For example, for normal incidence of plane waves, the scattering capability of the first order symmetric/anti-symmetric resonance is larger than that of the third order symmetric/anti-symmetric resonance, while the second order can hardly scatter any normally incident plane wave. This agrees with the transmission spectrum in Fig. 2, where only odd number order resonances are excited and the higher odd number order resonances are weaker for both symmetric and anti-symmetric modes. Secondly, the way of excitation/illumination also influences the field distribution and resonant frequency of a certain resonant mode. This is due to some retardation effects related to the finite structure size. For example, when we change the angle of incidence from $0°$ (i.e., normal incidence) toward $90°$, the first order anti-symmetric resonance would blue shift and the field distribution would also change slightly. Due to the retardation effect, the waveguide analysis can only give rough estimation of the resonant frequency of the sandwiched silver nanostrip.

Having demonstrated anti-symmetric localized surface plasmons on a sandwiched single silver nanostrip (2D case), we can further employ such anti-symmetric localized surface plasmons on a sandwiched single silver nanobrick (3D case) to operate as a magnetic metamaterial with optical magnetism. The inset in Fig. 4(a) shows an array of silver nanobricks sandwiched by SiC. The side length of each nanobrick (square shape) is 70 nm, the lattice constant is 120 nm, and the thicknesses of the silver layer and SiC layer are 20 nm and 30 nm, respectively. For normal incident light with the electric field polarized along the x direction, the transmission and reflection spectra are shown in Fig. 4(a). The effective permeability of the silver nanobrick array is retrieved from the transmission and reflection spectra [6], and is shown in Fig. 4(b). The Lorentz-type resonance in the dispersion spectra of the effective permeability confirms the optical magnetic response of the sandwiched single silver nanobrick. The transmission spectra are further given for different lattice constants and the magnetic resonances are nearly in the same spectra position as shown in Fig. 4(c). This verifies that the resonance is a local effect, where periodicity only influences the overall scattering amplitude. We further push the magnetic response to a shorter wavelength by reducing the thickness of the silver brick (Fig. 4(d)). Of course, the nanobricks are not limited to square shapes, other shapes, e.g., circular shapes, are also feasible, as long as the structure in the z direction is of IMI form.

In summary, we have demonstrated for the first time optical magnetic response in a single metal nanobrick. Our magnetic metamaterial allows a large magnetic response in the blue and violet light range, as well as in a part of the ultraviolet light range. This is also the first time a single metal layer is introduced to give large magnetic response. Our single-metal-layer design may stimulate new types of subwavelength magnetic resonators for applications in biosensing and optical imaging, etc. It also enriches the modes of localized surface plasmons, i.e., the interactions between the light and metal particles.

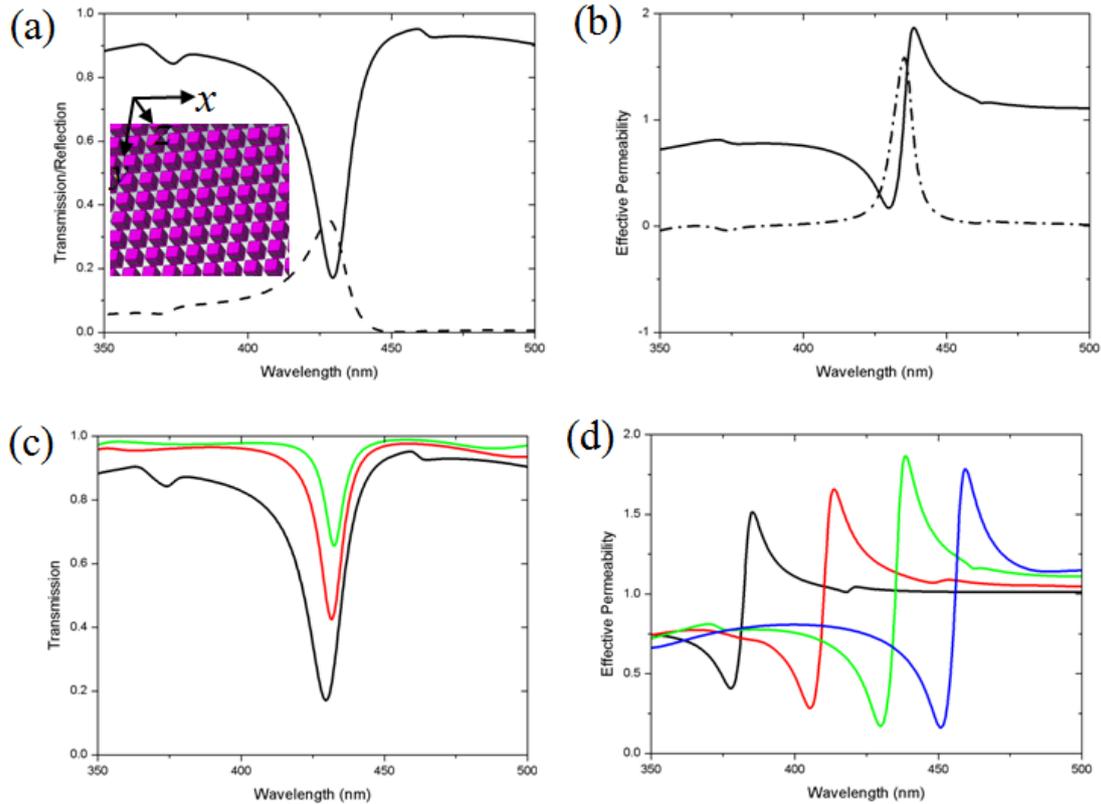

FIG. 4 (a) The transmission (solid line) and reflection (dashed line) spectra of an array of square silver nanobricks sandwiched by SiC. (b) The real (solid line) and imaginary (dash-dotted line) parts of the retrieved effective permeability. (c) The transmission spectra of the nanobrick samples with different lattice constants (120 nm for the black line, 200 nm for the red line, and 300 nm for the green line). (d) The effective permeability spectra of the nanobrick samples with different silver thickness (10 nm for the black line, 15 nm for the red line, 20 nm for the green line, and 25 nm for the blue line).

[1] A. N. Grigorenko, A. K. Geim, H. F. Gleeson, Y. Zhang, A. A. Firsov, I. Y. Khrushchev, and J. Petrovic, "Nanofabricated media with negative permeability at visible frequencies," Nature **438**, 335-338 (2005).

[2] V. M. Shalaev, W. S. Cai, U. K. Chettiar, H. K. Yuan, A. K. Sarychev, V. P. Drachev, and A. V. Kildishev, "Negative index of refraction in optical metamaterials," Optics Letters **30**, 3356-3358 (2005).

[3] Johnson, P. B. and R. W. Christy. "Optical Constants of the Noble Metals." Physical Review B 6, 4370 (1972).

[4] T. Sondergaard, and S. Bozhevolnyi, "Slow-plasmon resonant nanostructures: Scattering and field enhancements," Phys. Rev. B 75, 073402 (2007).

[5] E. N. Economou, "Surface Plasmons in Thin Films," Physical Review 182, 539 (1969).

[6] D. R. Smith, S. Schultz, P. Markos, and C. M. Soukoulis, "Determination of effective permittivity and permeability of metamaterials from reflection and transmission coefficients," Phys. Rev. B 65, 195104 (2002).